\documentclass[10pt, conference]{IEEEtran}
\IEEEoverridecommandlockouts

\usepackage[left=0.625in, right=0.625in, top=0.66in, bottom=0.95in, columnsep=0.12in]{geometry}

\usepackage{cite}
\usepackage{multicol}

\usepackage{amssymb}

\usepackage{amsmath}
\usepackage{amsfonts}
\usepackage[T1]{fontenc}
\usepackage{algpseudocode}
\usepackage[utf8]{inputenc}
\usepackage[english]{babel}
\usepackage[usenames, dvipsnames]{color}
\usepackage{pifont}
\usepackage{booktabs}
\usepackage{mathtools}
\usepackage{textcomp}
 \usepackage{graphicx}
\usepackage{subfigure}
\usepackage[ruled,vlined,commentsnumbered]{algorithm2e}
\usepackage{amsthm}
\usepackage{setspace}
\usepackage{url, hyperref}
\usepackage{enumitem}
\usepackage{nicefrac}

\SetKwInput{KwIn}{Inputs}

\theoremstyle{definition}

\theoremstyle{remark}

\theoremstyle{plain}

\newcommand{\RNum}[1]{\uppercase\expandafter{\romannumeral #1\relax}}

\def\BibTeX{{\rm B\kern-.05em{\sc i\kern-.025em b}\kern-.08em
    T\kern-.1667em\lower.7ex\hbox{E}\kern-.125em}}

\begin{document}

\title{Hierarchical Multi-Timescale Deep Reinforcement Learning
for QoS-Aware AI-RAN Configuration}
\author{
\IEEEauthorblockN{
Navid Keshtiarast\IEEEauthorrefmark{1},
Amirhassan Babazadeh Darabi\IEEEauthorrefmark{1},
Cem Tarhan\IEEEauthorrefmark{1},
Martin Stahn\IEEEauthorrefmark{2},\\
Ljiljana Simi\'c\IEEEauthorrefmark{1},
and Marina Petrova\IEEEauthorrefmark{1}
}
\IEEEauthorblockA{
\IEEEauthorrefmark{1}Chair of Mobile Communications and Computing, RWTH Aachen University, Germany; \IEEEauthorrefmark{2}T-Labs, Germany\\
Email: \{navid.keshtiarast, amirhassan.babazadeh, cem.tarhan, simic, petrova\}@mcc.rwth-aachen.de, martin.stahn@telekom.de
}
}
\maketitle

\begin{abstract}
Configuring radio access networks (RANs) for heterogeneous services requires coordinating radio parameters whose effects on throughput, delay, and reliability are closely coupled and evolve over different timescales. This paper presents a hierarchical deep reinforcement learning (DRL) framework for joint, QoS-aware RAN configuration that aligns the learning structure with the real-time, near-real-time, and non-real-time control loops of O-RAN that are also adopted by AI-RAN. Three proximal policy optimization agents jointly select per-user modulation and coding schemes, user-to-bandwidth-part (BWP) assignments, per-BWP transmit power, numerology, and BWP bandwidth weights, at three timescales. The agents share a QoS-aware objective combining throughput demand satisfaction with delay and reliability penalties. We extend ns-3 5G-LENA and integrate it with Ray RLlib through ns3-ai to support the required radio controls, network observations, and hierarchical policy execution.
We evaluate the agents' performance in a nine-cell downlink network with inter-cell interference, varying user populations, and mixed URLLC, gaming, conversational-video, and video-streaming traffic. Compared with the baseline 5G-NR RAN, our trained hierarchical DRL agents substantially reduce URLLC and conversational-video delays, lower the residual block error rate, and prevent the near-starvation of the weakest flows, trading off slightly lower throughput for high-rate video streaming under capacity-limited conditions.
\end{abstract}

\section{Introduction} \label{sec:Introduction}
The evolution toward 6G networks calls for increasingly efficient use of spectrum while accommodating diverse service requirements, from sustained high throughput for video streaming to stringent latency and reliability for interactive applications and industrial traffic~\cite{itu_imt2030}.
Conventional radio access network (RAN) resource allocation and configuration primarily aim to achieve high aggregate throughput, but this does not necessarily mean that delay and reliability requirements are satisfied for all services.
Maintaining a consistent user experience for all services therefore requires the RAN to adapt to changing traffic loads and channel conditions while accounting for service-specific QoS requirements~\cite{11141658}.


Such QoS-aware RAN adaptation is challenging due to two key factors. First, radio parameters are \textit{tightly coupled}~\cite{11007770}.
For example, a more robust modulation and coding scheme (MCS) can reduce decoding failures but uses more resources to deliver the same data. Increasing transmit power can improve the quality of one link while worsening interference elsewhere. Assigning users to different bandwidth parts (BWP) changes both their available capacity and, through the 5G-NR numerology, their transmission timescale. The appropriate choice for one parameter consequently depends on the others, as well as on the type of traffic being served. Second, these parameters inherently adapt at \textit{different timescales} in 3GPP networks. MCS must track the channel and HARQ dynamics within milliseconds; UE-to-BWP assignment and transmit power follow load and interference variations over tens of milliseconds; and numerology and BWP layout are carrier-level reconfigurations that are meaningful to adapt only over seconds~\cite{valcarce2024ai6gmac,polese2023oran}.

Learning-based control is a promising approach to this problem of agile RAN adaptation, as machine learning is expected to reshape 6G PHY and MAC functions such as resource allocation, adaptive modulation and coding, and power control~\cite{valcarce2024ai6gmac,nokia2024aiLayer2}. Within the context of the AI-native 6G RAN, the AI-RAN architecture provides the platform for such AI-for-RAN applications~\cite{airan2026arch}, based on the O-RAN control framework~\cite{11474793}. In O-RAN, control applications hosted by RAN intelligent controllers are organized by time scale: rApps handle non-real-time (non-RT) loops above 1~s; xApps handle near-RT loops between 10~ms and 1~s~\cite{polese2023oran}; and dApps are being introduced for RT loops below 10~ms~\cite{doro2022dapps}. The AI-RAN vision places learning-based control at each of these levels~\cite{11141658}.
The radio parameters discussed above fall naturally into these loops, with MCS in the RT loop, UE-to-BWP assignment and transmit power in the near-RT loop, and numerology and BWP layout in the non-RT loop. \textit{The key question then becomes how decisions taken in these different RAN control loops can jointly pursue common QoS objectives.}
A single flat agent cannot answer this question: if it acts at the fastest rate, it reconfigures structural parameters far too often and faces a combinatorial action space; acting at the slowest, it cannot follow the rate of channel variation. Conversely, controllers designed independently for each loop optimize local objectives and may conflict~\cite{11007770}.



\textit{Related work:} 
Existing work on learning-based RAN control typically addresses a single radio function or a single control loop~\cite{11370849}. At the RT level, ARIADNE~\cite{tsampazi2026ariadneairaninformedlink} applies reinforcement learning to MCS selection within a Sionna-based digital-twin environment. 
 At the near-RT level, Zhao et al.~\cite{Nakjung} propose AdaSlicing, which combines online learning with soft-isolated resource sharing to adapt slice allocations under changing network conditions while satisfying service-level requirements. Polese et al.~\cite{9814869} introduce ColO-RAN, where DRL-based xApps jointly select slice-level resource allocations and scheduling policies. These works either adapt a single function or coordinate a few decisions within one control loop, rather than jointly controlling coupled radio parameters across all control loops.
To the best of our knowledge, no prior work jointly learns link adaptation, BWP-level resource management, and carrier configuration across all three control timescales under a common QoS objective in a multi-cell network with inter-cell interference.

In this paper, we close this gap by \textit{aligning the learning controller with the timescale structure of the AI-RAN itself}. Three proximal policy optimization (PPO) agents adapt link-level parameters every 10~ms, BWP-level parameters every 50~ms, and carrier-level parameters on the scale of seconds, corresponding to the RT, near-RT, and non-RT loops of AI-RAN. All levels share one QoS-aware objective, so that slower decisions shape the operating point that faster decisions exploit. The contributions of this paper are summarized below. 

\begin{itemize}
    \item \textbf{Joint, QoS-aware RAN optimization problem across heterogeneous traffic classes:} We formulate RAN adaptation in a multi-cell network with inter-cell interference as a joint optimization problem of throughput demand satisfaction, delay, and reliability for heterogeneous traffic classes. The problem jointly configures coupled radio parameters, namely the per-UE MCS, UE-to-BWP assignment, per-BWP transmit power, numerology, and BWP widths, under class-specific QoS requirements. 

    \item \textbf{Hierarchical DRL framework for multi-timescale QoS-aware RAN control:} We decompose the coupled action space across three temporal resolutions that map to the RT, near-RT, and non-RT control loops of AI-RAN. We then learn all levels jointly with a hierarchy of three PPO agents under a common QoS-aware reward.


    \item \textbf{Full-stack system-level evaluation of the hierarchical DRL agents in ns-3 5G-LENA:} We integrate 5G-LENA with our ns3-ai/RLlib learning framework~\cite{Keshtiarast} through synchronized observation-action exchanges, executing each agent at its respective update period. To this end, we extend ns-3 5G-LENA so that the agents can control the per-UE MCS, the UE-to-BWP assignment, and the BWP numerologies and bandwidth weights at runtime. We evaluate our framework in a nine-cell network comprising three tri-sector sites with spectrum reuse and inter-cell interference. We consider randomized populations per cell, user positions, and traffic assignments to assess adaptation across different loads and mixtures of four traffic classes: URLLC, gaming, conversational video, and video streaming. Our results show that the proposed hierarchical DRL framework outperforms baseline 5G-NR RAN configurations in terms of delay for delay-sensitive traffic classes, residual BLER after HARQ, and throughput for several classes.

\end{itemize}

\section{System Model \& Problem Formulation}
\label{system_setup}
\subsection{System Model}
We consider a multi-cell downlink network with a set of gNBs $\mathcal G$, as illustrated in Fig.~\ref{fig:hdrl_model}. Each gNB $n\in\mathcal G$ comprises sectors $s\in\mathcal S_n$, where the pair $(n,s)$ identifies a cell. The complete cell set is  $J=\{(n,s):n\in\mathcal G,\ s\in\mathcal S_n\}$. 
The carrier bandwidth $W_{n,s}$ of cell $(n,s)$ is partitioned into a set of BWPs $\mathcal B_{n,s}$, where BWP $b$ has width $w_{n,s,b}$, numerology $\mu_{n,s,b}$, and transmit power $P_{n,s,b}$. 
 Let $\mathcal U_{n,s}$ denote the set of all UEs associated with cell $(n,s)$. Each UE carries exactly one downlink flow, so UEs and flows are in one-to-one correspondence. We denote by $\mathcal{K}$ the set of all flows and by $k$ the flow of UE $u$. Flow $k$ belongs to traffic class
$c(k) \in \mathcal{C}$.

Let $\mathbf x\in\mathcal X$ denote the joint network configuration, which collects all configured variables: 
\begin{equation}
\mathbf{x} = \big(\{m_u\},\
    \{a_{n,s,b,u}\},\{P_{n,s,b}\},\ \{\mu_{n,s,b}\},\ \{w_{n,s,b}\}\big),
  \label{eq:config}
\end{equation}
where $m_u \in \{0,\dots,28\}$ is the MCS index of UE $u$, and $a_{n,s,b,u} \in \{0,1\}$ equals one if UE $u \in \mathcal{U}_{n,s}$ is assigned to BWP $b$. $P_{n,s,b} \in [P_{\min}, P_{\max}]$ is the transmit power, $\mu_{n,s,b} \in \{0,1,2\}$ the numerology, and $w_{n,s,b}$ the width of BWP $b$ in cell $(n,s)$. The configuration space $\mathcal{X}$ is the Cartesian product of these per-UE and per-BWP ranges. Its dimension grows with the number of cells $|\mathcal{J}|$, BWPs per cell $B$, and UEs per cell $N_{\mathrm{UE}}$.

We model the traffic of cell $(n,s)$ as the aggregate of the flows of its UEs, where each flow generates packets according to the traffic profile of its class. Over an interval, let $A_k$ denote the newly arrived traffic of flow $k$, $Q_k^{\mathrm{start}}$ its queue backlog at the beginning of the interval, and $R_k$ the successfully served traffic~\cite{bolch2006queueing}. Further, let $d_k$ denote the mean packet delay of flow $k$ over the window, measured from packet arrival at the gNB to successful reception, and let $\widehat{e}_k$ denote its packet error rate. We consider a flow to be active if it contains newly arrived traffic, previously backlogged traffic, or both, so that the active flow set is $K^{\mathrm{act}}=\{k\in\mathcal K:A_k+Q_k^{\mathrm{start}}>0\}$.

The QoS profile of traffic class $c$ is defined as
\begin{equation}
\mathbf q_c=(\rho_c^{\min},D_c,\xi_c,E_c),
\label{eq:qos_profile}
\end{equation}
where $\rho_c^{\min}$ is the minimum acceptable throughput demand-satisfaction ratio, $D_c$ is the packet delay budget (PDB), $\xi_c$ is the maximum acceptable probability of exceeding this budget, and $E_c$ is the packet error rate (PER) requirement. The PDB and PER of each class follow its 5G QoS identifier (5QI)~\cite{3gpp_ts23501} and are listed in Table~\ref{tab:traffic_classes}. Each flow is evaluated jointly against its throughput, delay, and reliability requirements. The configuration $\mathbf{x}$ determines the outcome of every active flow: via its effects on resource availability, transmission efficiency, and inter-cell interference, it shapes the served traffic $R_k$, the delay $d_k$, and the packet error rate $\widehat{e}_k$ of flow $k$. For instance, the served traffic of flow $k$ of UE $u \in \mathcal{U}_{n,s}$ on BWP $b$ can be expressed as $R_k = f\big(\gamma_u,\ m_u,\ \mu_{n,s,b},\ w_{n,s,b},\ \ldots\big)$,
where $\gamma_u$ is the SINR of UE $u$, which depends on the transmit power $P_{n,s,b}$ and on the inter-cell interference from cells that reuse the same carrier.


\subsection{QoS Objectives}
\label{QoS_Objectives}

To evaluate a RAN configuration $\mathbf{x}$ against the per-class requirements $q_{c(k)}$, we define one objective for each QoS requirement, i.e. throughput-demand satisfaction, delay-budget satisfaction, and reliability-budget satisfaction, each averaged over the active flows.

As the throughput QoS metric, we define the throughput demand-satisfaction ratio of flow $k$ as $\rho_k = \frac{R_k}{A_k + Q_k^{\mathrm{start}} + \epsilon}$,
where $\epsilon > 0$ is a small constant that prevents division by zero. $\rho_k$ is the per-flow quantity that is compared with the class threshold $\rho_{c(k)}^{\min}$ in~\eqref{eq:qos_profile}.
It approaches one when the flow's newly arrived and previously backlogged traffic is fully served. The first objective maximizes throughput demand satisfaction:
\begin{equation}
  f_1(\mathbf{x}) = \frac{1}{|\mathcal{K}^{\mathrm{act}}|}
  \sum_{k\in\mathcal{K}^{\mathrm{act}}} \min(\rho_k, 1).
  \label{eq:demand_satisfaction}
\end{equation}


The second objective maximizes delay-budget satisfaction:
\begin{equation}
f_2(\mathbf x)=1-\frac{1}{|\mathcal K^{\mathrm{act}}|}\sum_{k\in\mathcal K^{\mathrm{act}}}\min(\delta_k,1),
\label{eq:delay_objective}
\end{equation}
with normalized delay overshoot $\delta_k=\text{max}(\frac{d_k-D_{c(k)}}{D_{c(k)}},0)$.
The delay penalty is zero when the measured delay $d_k$ satisfies the class-specific budget and increases with the relative magnitude of a violation.

The third objective maximizes reliability-budget satisfaction:
\begin{equation}
f_3(\mathbf x)=1-\frac{1}{|\mathcal K^{\mathrm{act}}|}\sum_{k\in\mathcal K^{\mathrm{act}}}\min(\varepsilon_k,1),
\label{eq:reliability_objective}
\end{equation}
where $\varepsilon_k= \text{max}(\left[\log_{10}(\widehat e_k+\epsilon)-\log_{10}E_{c(k)}\right],0)$.
The logarithmic overshoot expresses reliability violations relative to the class-specific target across different orders of magnitude.

\subsection{Joint QoS-Aware RAN Optimization Problem}
The joint RAN configuration problem is formulated as:
\begin{subequations}
\label{eq:optimization}
\begin{align}
\underset{\mathbf x\in\mathcal X}{\operatorname{max.}}\quad &\mathbf F(\mathbf x)=\bigl(f_1(\mathbf x),f_2(\mathbf x),f_3(\mathbf x)\bigr)\\
\text{s.t.}
\quad &\rho_k \ge \rho_{c(k)}^{\min}\quad \forall k \in \mathcal{K}^{\mathrm{act}},\label{eq:min_demand}\\
&\Pr[d_k>D_{c(k)}]\leq\xi_{c(k)},\quad\forall k\in\mathcal K^{\mathrm{act}},\label{eq:delay_constraint}\\
&\widehat e_k\leq E_{c(k)},\quad\forall k\in\mathcal K^{\mathrm{act}},\label{eq:error_constraint}\\
&\lim_{T\to\infty}\frac{\mathbb E[Q_k(T)]}{T}=0,\quad\forall k\in\mathcal K,\label{eq:queue_stability}\\
&\sum_{b\in\mathcal B_{n,s}}w_{n,s,b}\leq W_{n,s},\quad\forall(n,s)\in\mathcal J,\label{eq:bandwidth_constraint}\\
&\sum_{b\in\mathcal B_{n,s}}a_{n,s,b,u}=1,\;\forall u\in\mathcal U_{n,s},\, \forall(n,s)\in\mathcal J.\label{eq:assignment_constraint}
\end{align}
\end{subequations}
Constraints~\eqref{eq:min_demand} to \eqref{eq:error_constraint} specify the throughput, delay, and reliability requirements. Constraint~\eqref{eq:queue_stability} imposes mean-rate queue stability, while \eqref{eq:bandwidth_constraint} and \eqref{eq:assignment_constraint} enforce bandwidth feasibility and a single serving BWP per UE.


\section{Hierarchical DRL Framework for QoS-Aware RAN Configuration}
\label{sec:actions}
\begin{figure*}
    \centering
    \includegraphics[width=0.9\linewidth,trim = 12mm 17mm 12mm 37mm,clip]{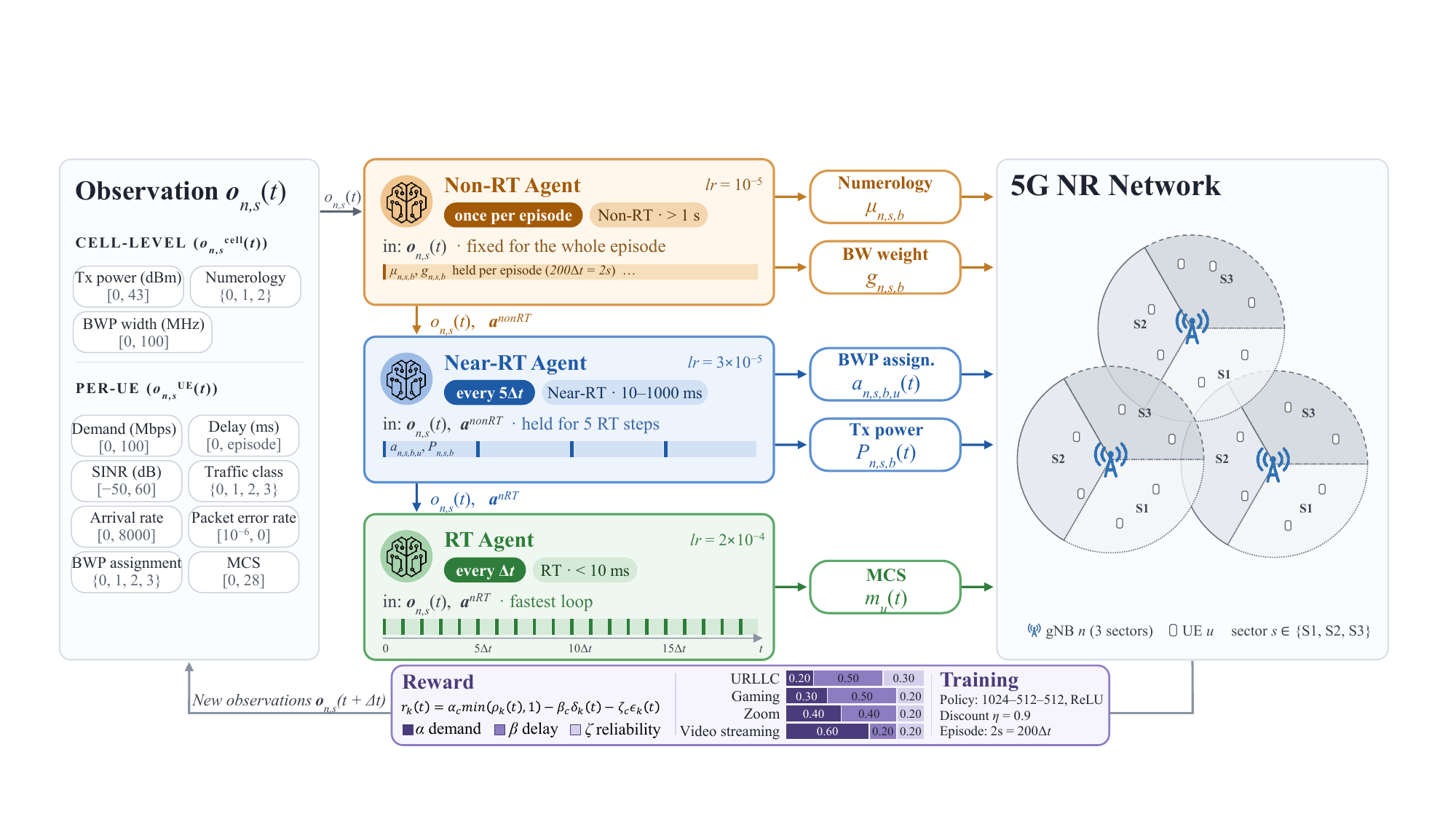}
    \vspace{-2mm}
    \caption{Hierarchical DRL framework for QoS-aware AI-RAN configuration. Evaluation scenario is a 9-cell network of 3 tri-sector gNB sites and 5-15 UEs per cell; sector color indicates cell's carrier frequency.}
    \label{fig:hdrl_model}
         \vspace{-4mm}
\end{figure*}

Problem~\eqref{eq:optimization} is a constrained multi-objective problem over a high-dimensional, mixed discrete and continuous configuration space whose parameters are adapted on different timescales in practical RANs.
We therefore address the problem with a hierarchical DRL framework, as illustrated in Fig.~\ref{fig:hdrl_model}. 
Since the mapping from $\mathbf{x}$ to the QoS in~\eqref{eq:optimization} has no closed form, we model the problem as a Markov decision process (MDP) $(\mathcal{S}, \mathcal{A}, \mathcal{P}, r, \eta)$, where the state $s(t) = \mathbf{o}(t)$ is the network observation (Section~\ref{sec:observations}), the action $\mathbf{a}(t)$ realizes $\mathbf{x}$ (Section~\ref{sec:actions}), $r$ is the reward (Section~\ref{sec:reward}), and $\eta$ is the discount factor. Each agent $i \in \{\mathrm{RT}, \mathrm{nRT}, \mathrm{nonRT}\}$ learns a policy $\pi_i(\mathbf{a}^i \,|\, s(t))$ with proximal policy optimization (PPO)~\cite{schulman2017proximal}.
\subsection{Hierarchical Action Space}
\label{sec:actions}

\begin{table}[t]
\centering
\footnotesize
\caption{Traffic classes \& corresponding QoS requirements.}
\vspace{-2mm}
\label{tab:traffic_classes}
\renewcommand{\arraystretch}{1.4}
\setlength{\tabcolsep}{5pt}

\begin{tabular}{|p{0.20cm}|p{1.95cm}|p{0.6cm}|p{0.8cm}|p{3.6cm}|}
\hline
\textbf{C} &
\textbf{Traffic class} &
\textbf{PER} &
 \textbf{PDB}&
\textbf{Traffic profile} \\
\hline

1 &
URLLC \cite{etsi38913} &

$10^{-5}$ &
1 ms &
Packet size: 32--100 B\newline periodic arrival interval: 1 ms
\\
\hline

2 &
DL gaming \cite{bojovic2022ngmn,3gpp_ts23501} &

$10^{-3}$ &
50 ms &
packet arrival interval: 58.5 ms
\newline
packet size: 140.8 B
\\
\hline

3 &
Conversational video \cite{zoom_bandwidth,3gpp_ts23501} &

$10^{-3}$ &
150 ms&
[HQ, 720p HD, 1080p HD]: [0.6, 1.2, 3] Mbit/s.
\\
\hline

4 &
Video streaming \cite{netflix_bandwidth,3gpp_ts23501} &
$10^{-6}$ &
300 ms &
[720p HD, 1080p HD, 4K]: [3, 4.8, 15] Mbit/s
\\
\hline

\end{tabular}
     \vspace{-3mm}
\end{table}

To account for the different timescales of the parameters in $\mathbf x$, we divide the joint action space among three agents: an RT agent, a near-RT agent, and a non-RT agent, corresponding to the respective control loops of AI-RAN. 
Let $\Delta t=10~\text{ms}$ denote the basic control interval. The RT agent acts every interval, the near-RT agent acts every five intervals, i.e., every 50\,ms, and the non-RT agent acts once at the beginning of each episode, which spans 200 intervals, i.e., 2\,s.

\subsubsection{RT Agent}
The RT Agent selects the MCS of each UE:
\begin{equation}
\mathbf a^{\mathrm RT}(t)=\{m_{u}(t)\}_{u\in\mathcal{U}}.
\end{equation}
These decisions replace the default 5G-NR CQI-based link adaptation and adapt the transmission configuration to the current radio conditions, influencing the achievable service rate and decoding reliability.
\subsubsection{Near-RT Agent}
The near-RT agent jointly selects the serving BWP of each UE and the transmit power of each cell's BWPs:
\begin{equation}
\mathbf a^{\mathrm nRT}(t)=\left(\{a_{n,s,b,u}(t)\}_{(n,s),b,u},\{P_{n,s,b}(t)\}_{(n,s),b}\right).
\end{equation}
The assignment variables satisfy the single-BWP constraint~\eqref{eq:assignment_constraint}. In the implementation, the agent selects one BWP index per UE, which directly defines this assignment.
The transmit power satisfies the minimum and maximum constraints
$P_{n,s,b}(t) \in \{33,\dots,43\}~\text{dBm}.$
BWP assignment determines the resources and numerology available to each UE, while transmit power affects both the desired signal strength and the interference experienced by neighboring cells. These decisions remain unchanged between consecutive near-RT agent updates.

\subsubsection{Non-RT Agent}
The non-RT Agent selects the numerology and bandwidth weight of each BWP in each cell:
\begin{equation}
\mathbf a^{\mathrm nonRT}=\big(\{\mu_{n,s,b}\}_{(n,s),b}, \{g_{n,s,b}\}_{(n,s),b}\big),
\end{equation}
where $g_{n,s,b} \in \{1,\ldots,10\}$ is bandwidth weight and sets the size of BWP $b$, i.e., the cell bandwidth is divided among the BWPs in proportion to their weights,
$w_{n,s,b}=W_{n,s}\, \frac{g_{n,s,b}}{\sum_{b'\in\mathcal{B}_{n,s}} g_{n,s,b'}}$.
The non-RT agent's decisions establish the BWP layout at the start of each episode and remain fixed throughout that episode.


The complete network action combines the latest decisions of all three agents:
\begin{equation}
  \mathbf{a}(t) = \big(\mathbf{a}^{\mathrm{RT}},\
  \mathbf{a}^{\mathrm{nRT}},\
  \mathbf{a}^{\mathrm{nonRT}}\big),
\end{equation}
Together, these actions realize the configuration $\mathbf x$ in~\eqref{eq:config}.
This decomposition preserves joint control while allowing each parameter group to be adapted at its assigned timescale.
\subsection{Observation Space}
\label{sec:observations}
The observation space provides the agents with the current network configuration, i.e., the transmit power, numerology, and width of each BWP, radio conditions, i.e., the SINR $\gamma_u$ of each
UE, traffic characteristics, and the QoS measurements $R_k$, $d_k$, and $\widehat{e}_k$ of each flow. For each cell $(n,s)$, the observation consists of the cell-level observation $\mathbf{o}^{\mathrm{cell}}_{n,s}(t)$, and per-UE observations $\mathbf{o}^{\mathrm{UE}}_u(t)$ of its associated UEs:
\begin{equation}
\mathbf o_{n,s}(t)=\left[\mathbf o_{n,s}^{\mathrm{cell}}(t),\{\mathbf o_u^{\mathrm{UE}}(t)\}_{u\in\mathcal U_{n,s}}\right].
\end{equation}
The network observation concatenates the observations of all cells, $\mathbf{o}(t) = \big[\mathbf{o}_{n,s}(t)\big]_{(n,s)\in\mathcal{J}}.$
Fig.~\ref{fig:hdrl_model} lists the observed fields and their ranges. The cell-level fields describe the current BWP configuration. The per-UE fields describe the radio conditions, the offered traffic, the applied control decisions, and the achieved QoS. Throughput, delay, and packet error rate correspond to the quantities $R_k$, $d_k$, and $\widehat e_k$ used in the objectives of Section~\ref{system_setup}-B, which allows the agents to relate their decisions to the resulting QoS. 
Since the number of UEs per cell varies across episodes, the per-UE observations are zero-padded to a fixed size.
The three agents differ in the temporal resolution of their observations.
Network measurements are collected at the basic control interval $\Delta t$, which is the observation interval of the RT agent. The near-RT agent observes measurements averaged over its decision interval of $5\Delta t$. At episode initialization, performance measurements for the new episode are not yet available, so the non-RT agent observes the initial configuration and scenario information.
\subsection{Reward Function}
\label{sec:reward}
We combine the three QoS objectives in~\eqref{eq:optimization} into a scalar reward to guide DRL-based optimization. The QoS requirement constraints~\eqref{eq:min_demand} to \eqref{eq:error_constraint} are encoded as penalties rather than enforced explicitly, while constraints~\eqref{eq:bandwidth_constraint} and~\eqref{eq:assignment_constraint} are satisfied by construction of the action space in Section~\ref{sec:actions}. For flow $k$, i.e., the single flow carried by UE $u$ with traffic class $c=c(k)$, the reward is
\begin{equation}
r_k(t)=\alpha_c \min\big(\rho_k(t),1\big) - \beta_c\,\delta_k(t) - \zeta_c\,\varepsilon_k(t),
\label{eq:ue_reward}
\end{equation}
where $\rho_k(t)$, $\delta_k(t)$, and $\varepsilon_k(t)$ are the per-flow terms as defined in Section~\ref{QoS_Objectives}. 
The demand-satisfaction term rewards the fraction of newly arrived and backlogged traffic that is
successfully served. The delay and reliability terms penalize violations of the corresponding class-specific budgets.
The traffic class-dependent weights $\alpha_c$, $\beta_c$, and $\zeta_c$ control the relative importance of these components. We set these weights according to the relative QoS requirements of the considered traffic classes, as given by their standardized 5G QI characteristics in 3GPP TS~23.501~\cite{3gpp_ts23501}.  They allow the agents to learn different trade-offs according to the throughput, latency, and reliability requirements of each traffic class.

The cell reward $r_{n,s}(t)$ is the average of $r_k(t)$ over the active flows of cell $(n,s)$, and the network reward $r_{\mathrm{net}}(t)$ is the average of $r_{n,s}(t)$ over all cells with at least one active flow. All three agents share $r_{\mathrm{net}}(t)$, averaged over their respective decision periods, i.e., over $\Delta t$, $5\Delta t$, and one episode for the RT, near-RT, and non-RT agents, respectively.



\section{Performance Evaluation}
\label{sec:evaluation}

\subsection{Simulation \& Training Setup}
Our simulation framework integrates the ns-3 network simulator with Ray RLlib~\cite{liang2018rllib} through ns3-ai~\cite{ns-3ai, Keshtiarast}. The 5G-LENA module~\cite{Patriciello2019AnES} models the 5G-NR network and its PHY and MAC functionalities, while Ray RLlib implements PPO-based agent training. 
At each time step, ns-3 collects the current network configuration, radio measurements, traffic information, and achieved QoS. These observations are transferred to the learning environment, which obtains the decisions of the agents scheduled to act at that instant. The resulting configuration combines the latest decisions of the three agents and is applied to the simulator. The simulation then advances to the next observation interval, and the resulting performance measurements are used to compute the reward. We run 50 independent workers, each running one simulation instance in parallel, to collect training experience under different network realizations.


Our simulation scenario in Fig.~\ref{fig:hdrl_model} comprises three gNB sites with three sectors per site, i.e., nine cells, operating at  3.3~GHz with 20~MHz of bandwidth per cell. 
Propagation is modeled using the  3GPP Urban Macro model with line-of-sight and non-line-of-sight conditions and small-scale channel variations, and the gNBs use the 3GPP antenna model. At the beginning of each training and evaluation episode, the number of UEs in each cell is randomly selected between $5$ and $15$, giving a total population of 45 to 180 UEs, and each UE is placed randomly and assigned one of the four traffic classes in Table~\ref{tab:traffic_classes}.  This exposes the agents to different radio conditions, interference levels, and offered loads. Because the agents require a fixed input size, we zero-pad per-UE observations for cells with fewer than 15 UEs.
The traffic mix combines small, delay-sensitive flows with high-rate video. URLLC has the strictest delay budget of 1~ms, whereas buffered video streaming has the most relaxed budget of 300~ms but generates the highest load of up to 15~Mbit/s per flow. Fig.~\ref{fig:hdrl_model} summarizes the simulation and training parameters. 

\subsection{Baselines \& Evaluation Metrics}
We compare our hierarchical DRL agents against baseline 5G-NR configurations with round-robin (RR), proportional-fair (PF), and QoS-aware scheduling. In these baseline configurations, MCS selection uses a default 5G-NR CQI-based approach. The baseline radio configuration uses one BWP, numerology $\mu=1$, and a fixed transmit power of 43~dBm. Our hierarchical DRL agents operate on top of the QoS-aware scheduler, so the QoS baseline isolates the effect of the learned configuration. All configurations are evaluated on the same 50 episodes, each with an independent UE population, UE positions, and traffic assignment. 
We assess the proposed DRL and baseline 5G-NR configurations based on three key metrics: throughput, packet delay, and reliability measured by the residual BLER after HARQ. Since a packet is lost if any of the transport blocks carrying it fails after HARQ, the packet error rate follows from the residual BLER $r$ as $\mathrm{PER} = 1-(1-r)^{N_{\mathrm{eff}}}$; as the considered packets are small compared with the transport block size, $N_{\mathrm{eff}} \approx 1$ and the residual BLER is directly compared with the 5QI PER targets in Table~\ref{tab:traffic_classes}. 

\subsection{Results}
\begin{figure}[t]
  \centering
  \includegraphics[width=0.95\columnwidth]{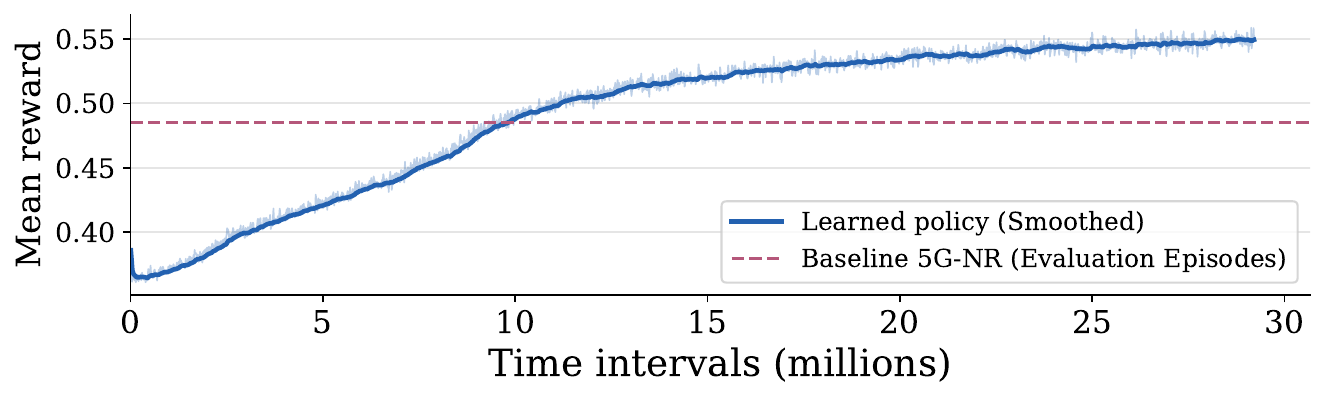}
  \vspace{-4mm}
  \caption{Training reward over time intervals.}
  \label{fig:training}
  \vspace{-4.5mm}
\end{figure}
First, we evaluate the training and convergence of our hierarchical DRL agents. Fig.~\ref{fig:training} shows the evolution of the mean network reward per basic interval over 29 million intervals. After an initial drop, the smoothed reward increases steadily from approximately $0.37$ to $0.55$, exceeding the reward of the QoS baseline (on the evaluation episodes) after 10 million steps (time intervals). The rate of improvement subsequently decreases, and the curve gradually converges toward the end of training.

\begin{figure}[!t]
  \centering
  \setlength{\subfigcapskip}{-6.5pt}
  \includegraphics[width=0.48\textwidth]{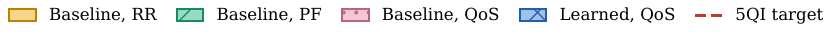}\\[-1mm]
  \subfigure[Delay]{
    \includegraphics[width=0.80\columnwidth]{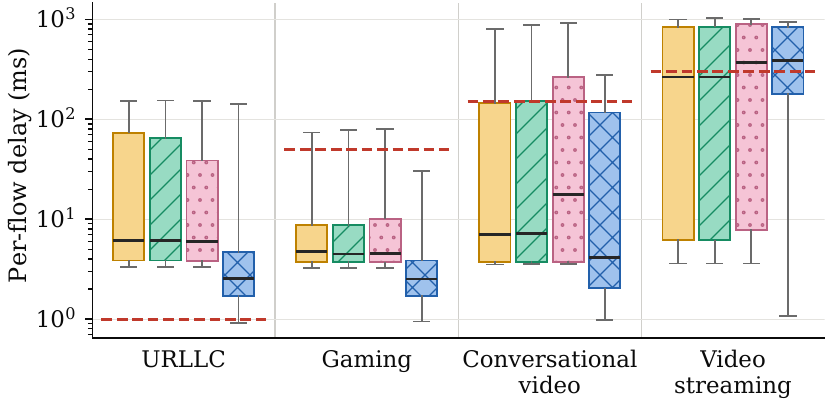}
    \label{TSUS_1_1}
  }\\[-1.5mm]
  \subfigure[Residual BLER]{
    \includegraphics[width=0.80\columnwidth]{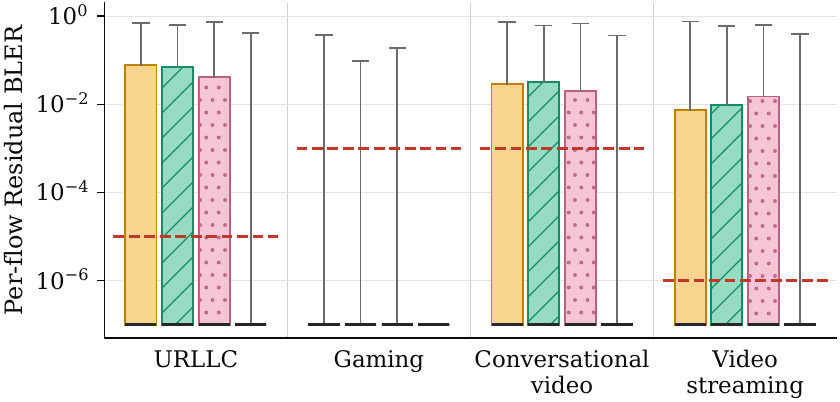}
    \label{TSUS_1_2}
  }\\[-1.5mm]
  \subfigure[Throughput]{
    \includegraphics[width=0.80\columnwidth]{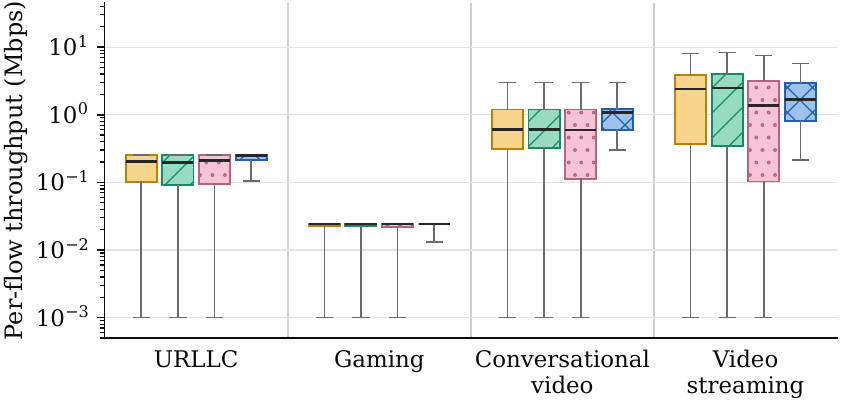}
    \label{TSUS_1_3}
  }\\[-3mm]
  \caption{Per-flow delay, residual BLER after HARQ, and throughput for each traffic class under our hierarchical DRL agents and the 5G-NR baseline configurations. Dashed lines indicate the 5QI targets per traffic class.}
  \label{fig:results_6nodes}
  \vspace{-0.4cm}
\end{figure}

Fig.~\ref{fig:results_6nodes} presents the resulting per-class QoS performance of our hierarchical DRL framework versus the 5G-NR baselines. Fig.~\ref{TSUS_1_1} shows the distributions of per-flow mean packet delay for the four traffic classes. Our hierarchical DRL agents reduce the median URLLC delay from about 6~ms under all three 5G-NR baselines to about 2.7~ms, with a narrower interquartile range indicating more consistent performance across flows. This is because the agents use the BWPs more efficiently by adapting their sizes and numerologies to the served traffic. Our lowest achieved URLLC delay reaches the stringent 1~ms budget, although the typical delay remains above this threshold. For gaming, all configurations satisfy the 50~ms budget, and our hierarchical DRL agents further reduce the median delay from about 4.5~ms to about 2.5~ms. For conversational video, our hierarchical DRL agents reduce the median delay to about 4~ms, compared with about 7~ms for RR and PF and about 17~ms for the QoS baseline. Our upper quartile delay stays below the 150~ms budget, though the upper whisker still extends beyond the budget, indicating that a few flows continue to experience excessive delays. By contrast, for video streaming our hierarchical DRL agents overall yield a higher delay than the 5G-NR baselines, with a median delay above the 300~ms budget. In heavily loaded episodes with many users and high-rate video streams, the offered traffic can exceed the capacity available within the considered 20~MHz bandwidth per cell, making it difficult to satisfy all flows simultaneously. Under these conditions, our hierarchical DRL agents prioritize the latency requirements of delay-sensitive services over the demands of high-throughput streaming flows. 

Fig.~\ref{TSUS_1_2} shows the distribution of the residual BLER after HARQ retransmissions. 
The 5G-NR baselines exhibit wide distributions for URLLC, conversational video, and video streaming, with interquartile ranges extending well above the corresponding 5G target. In contrast, the interquartile range of our hierarchical DRL agents lies at the lower plotting limit for all classes. This is because the agents select lower, more robust MCS values for the
small packets of URLLC and gaming and adapt the transmit power.

Fig.~\ref{TSUS_1_3} presents the distribution of per-flow mean throughput. For the 5G-NR baselines, the weakest flows of every class drop to about $10^{-3}$~Mbit/s and are nearly starved. Our hierarchical DRL agents avoid such starvation and increase the lower whisker (at 1.5 times the interquartile range from the median) throughput of every class by at least an order of magnitude. For URLLC and gaming, which carry low offered rates, the median throughput is similar across configurations, showing that the delay and reliability gains of our hierarchical DRL agents come without a throughput loss for these classes.
For conversational video, our hierarchical DRL agents increase the median throughput from about 0.6 to about 1~Mbit/s and yield a narrower distribution than all baselines. For video streaming, our median throughput is lower than that of RR and PF but higher than that of the QoS 5G-NR baseline. Together with the delay results in Fig.~\ref{TSUS_1_1}, Fig.~\ref{TSUS_1_3} thus reveals a slight service trade-off: improved performance for delay-sensitive traffic is accompanied by somewhat reduced typical throughput for high-demand streaming flows. Such an outcome is consistent with our reward design, which evaluates demand satisfaction and class-dependent delay and reliability penalties rather than maximizing the total number of delivered bits alone. 
\section{Conclusions}\label{sec:Conclusions}
This paper presented a hierarchical DRL approach that jointly adapts coupled RAN parameters across multiple timescales to serve heterogeneous QoS requirements. By aligning three PPO agents with the RT, near-RT, and non-RT control loops of AI-RAN, the approach adapts the MCS, the UE-to-BWP assignment and transmit power, and the BWP layout at the timescale on which each parameter is adapted. Using ns-3, we trained and evaluated the agents in a nine-cell network with inter-cell interference and varying traffic loads. Compared with baseline 5G-NR configurations with RR, PF, and QoS-aware scheduling, our agents reduced the median delay of URLLC, gaming, and conversational video, lowered the residual BLER after HARQ and prevented the near-starvation of the weakest flows at the cost of higher delay and lower throughput for high-rate video streaming under capacity-limited conditions. 
Future work will investigate distributed learning to support local decisions and coordination among neighboring cells.
\bibliographystyle{IEEEtran}
\bibliography{bib}

@ARTICLE{11370849,
  author={Kirana, Mira Chandra and others},
  journal={IEEE Commun. Surveys Tuts.}, 
  title={ML-Enabled Open RAN: A Comprehensive Survey of Architectures, Challenges, and Opportunities}, 
  year={2026},
  volume={28},
  number={},
  pages={4948-4979}}

@techreport{airan2026arch,
  author      = {{AI-RAN Alliance}},
  title       = {{AI-RAN} Architecture Overview and Component Definitions},
  institution = {AI-RAN Alliance},
  type        = {Technical Report},
  number      = {Version 1.2},
  year        = {2026},
  url         = {https://ai-ran.org}
}

@book{bolch2006queueing,
  author    = {Gunter Bolch and others},
  title     = {Queueing Networks and Markov Chains:
               Modeling and Performance Evaluation
               with Computer Science Applications},
  edition   = {2},
  publisher = {John Wiley \& Sons},
  year      = {2006},
  doi       = {10.1002/0471791571},
  isbn      = {978-0-471-56525-3}
}

@article{polese2023oran,
  author={Polese, Michele and others},
  journal={IEEE Communications Surveys \& Tutorials}, 
  title={{Understanding O-RAN: Architecture, Interfaces, Algorithms, Security, and Research Challenges}}, 
  year={2023},
  volume={25},
  number={2},
  pages={1376-1411}}

@article{doro2022dapps,
   author={D'Oro, Salvatore and others},
  journal={IEEE Commu. Mag.}, 
  title={{dApps: Distributed Applications for Real-Time Inference and Control in O-RAN}}, 
  year={2022},
  volume={60},
  number={11},
  pages={52-58}}

@ARTICLE{11007770,
  author={Brach del Prever, Pietro and others},
  journal={IEEE Transactions on Mobile Computing}, 
  title={{PACIFISTA: Conflict Evaluation and Management in Open RAN}}, 
  year={2025},
  volume={24},
  number={10},
  pages={10590-10605},
  doi={10.1109/TMC.2025.3570632}}

@inproceedings{valcarce2024ai6gmac,
  author    = {Valcarce, Alvaro and Kela, Petteri and
               Mandelli, Silvio and Viswanathan, Harish},
  title     = {The Role of {AI} in {6G} {MAC}},
  booktitle = {Proc. EuCNC/6G Summit},
  year      = {2024},
  pages     = {723--728},
}

@techreport{nokia2024aiLayer2,
  author      = {{Nokia}},
  title       = {{AI} Opportunities in {6G} Layer 2},
  institution = {Nokia},
  type        = {White paper},
  number      = {},
  year        = {2024},
  month       = apr,
  url         = {https://www.nokia.com/asset/213703/}
}

@techreport{itu_imt2030,
  author      = {{ITU-R}},
  title       = {Framework and Overall Objectives of the Future
                 Development of {IMT} for 2030 and Beyond},
  institution = {ITU},
  type={Recommendation},
  number      = {M.2160-0},
  year        = {2023}
}

@ARTICLE{11141658,
  author={Kundu, Lopamudra and others},
  journal={IEEE Commun. Mag.}, 
  title={{AI-RAN: Transforming RAN with AI-Driven Computing Infrastructure}}, 
  year={2026},
  volume={64},
  number={1},
  pages={168-174},
  doi={10.1109/MCOM.001.2500018}}

@ARTICLE{11474793,
  author={Polese, Michele and others},
  journal={IEEE Commun. Mag.}, 
  title={{Beyond Connectivity: An Open Architecture for AI-RAN Convergence in 6G}}, 
  year={2026},
  volume={64},
  number={7},
  pages={90-95},
 }

@techreport{etsi38913,
  author      = {{ETSI}},
  title       = {{5G; Study on Scenarios and Requirements for Next Generation Access Technologies}},
  institution = {ETSI},
  type        = {TR},
  number      = {138 913 V16.0.0},
  year        = {2020}
}

@techreport{3gpp_ts23501,
  author      = {{3GPP}},
  title       = {{System Architecture for the {5G} System ({5GS})}},
  institution = {3GPP},
  type        = {TS},
  number      = {3GPP TS 23.501 V16.7.0},
  year        = {2021},
}

@inproceedings{bojovic2022ngmn,
  author    = {Bojovic, Biljana and Lag{\'e}n, Sandra},
  title     = {{Enabling NGMN Mixed Traffic Models for ns-3}},
  booktitle = {Proc. WNS3},
  year      = {2022},
  pages     = {127--134}
}

@misc{zoom_bandwidth,
  author       = {{Zoom}},
  title        = {{Zoom System Requirements: Windows, macOS, Linux}},
  howpublished = {Zoom Support},
  year         = {2026},
  note         = {Accessed: 2026\-08\-11}
}

@misc{netflix_bandwidth,
  author       = {{Netflix}},
  title        = {Netflix-Recommended Internet Speeds},
  howpublished = {Netflix Help Center},
  year         = {2026},
  url          = {{https://help.netflix.com/en/node/306}}
}

@ARTICLE{Keshtiarast,
  author={Keshtiarast, Navid and Renaldi, Oliver and Petrova, Marina},
  journal={IEEE Networking Letters}, 
  title={{Wireless MAC Protocol Synthesis and Optimization With Multi-Agent Distributed Reinforcement Learning}}, 
  year={2024},
  volume={6},
  number={4},
  pages={242-246}}

@article{Patriciello2019AnES,
author = {Natale Patriciello and others},
title = {{An E2E simulator for 5G NR networks}},
journal = {Simul. Model. Pract. and Theory},
volume = {96},
pages = {101933},
year = {2019},
}

@INPROCEEDINGS{Nakjung,
  author={Zhao, Ming and others},
  booktitle={Proc. IEEE INFOCOM}, 
  title={AdaSlicing: Adaptive Online Network Slicing Under Continual Network Dynamics in Open Radio Access Networks}, 
  year={2025},
  volume={},
  number={},
  pages={1-10}}

@misc{tsampazi2026ariadneairaninformedlink,
      title={{ARIADNE: AI-RAN Informed Link Adaptation in Digital Twin Network Environments}}, 
      author={Maria Tsampazi and others},
      year={2026},
      eprint={2605.29772},
      archivePrefix={arXiv},
      primaryClass={cs.NI},
      url={https://arxiv.org/abs/2605.29772}, 
}

@ARTICLE{9814869,
  author={Polese, Michele and others},
  journal={IEEE Trans. on Mobile Comput.}, 
  title={{ColO-RAN: Developing Machine Learning-Based xApps for Open RAN Closed-Loop Control on Programmable Experimental Platforms}}, 
  year={2023},
  volume={22},
  number={10},
  pages={5787-5800}}

@inproceedings{liang2018rllib,
    title={{RLlib: Abstractions for Distributed Reinforcement Learning}},
    author={
        Liang, Eric and others
    },
    booktitle = {Proc. ICML},
    year={2018}
}

@article{schulman2017proximal,
      title={{Proximal Policy Optimization Algorithms}}, 
      author={John Schulman and others},
      journal={arXiv preprint arXiv:1707.06347},
      year={2017},
}

@inproceedings{ns-3ai,
author = {Yin, Hao and others},
title = {{Ns3-Ai: Fostering Artificial Intelligence Algorithms for Networking Research}},
booktitle = {Proc. WNS3},
year = {2020},
pages     = {57--64}
}

@misc{delay,
  note = {Accessed date: 2025-05-28. [Online]. Available: \url{http://www.wirelesscommunication.nl/reference/chaptr03/fading/delayspr.htm}},
  title = {Delay Spread}
}
\end{document}